\documentclass[acmsmall,screen,nonacm]{acmart}

\authorsaddresses{}

\title{\tool: Causal Black-Box Testing of Machine Unlearning}

\author{Anna Mazhar}
\affiliation{%
  \institution{Cornell University}
  \city{Ithaca}
  \state{NY}
  \country{USA}}
\email{annam@cs.cornell.edu}

\author{Sainyam Galhotra}
\affiliation{%
  \institution{Cornell University}
  \city{Ithaca}
  \state{NY}
  \country{USA}}
\email{sg@cs.cornell.edu}

\usepackage{tabularx}
\usepackage{booktabs}
\usepackage{hyperref}
\usepackage{multirow}
\usepackage{listings}

\usepackage{makecell}
\usepackage{graphicx}
\usepackage{subcaption}
\usepackage{wrapfig}
\usepackage{framed}
\usepackage{amsthm}
\usepackage{xcolor}
\usepackage{flushend}
\usepackage{pifont} %
\usepackage{amsmath}
\usepackage{paralist}
\usepackage{siunitx}
\usepackage{enumitem}
\usepackage{algorithm}
\usepackage[noend]{algorithmic}
\usepackage{fancyhdr}
\usepackage{tikz}
\usetikzlibrary{arrows.meta,positioning,calc,fit,backgrounds,shapes.geometric}

\definecolor{cafeaccent}{HTML}{0B5FA5}
\definecolor{cafecausal}{HTML}{C1662F}
\definecolor{cafetruth}{HTML}{111827}
\definecolor{cafemuted}{HTML}{6B7280}
\definecolor{cafefaint}{HTML}{B9BEC7}
\definecolor{cafegrid}{HTML}{D6DAE0}

\tikzset{
  cnode/.style={draw=cafemuted, fill=white, rounded corners=2pt,
                inner sep=2.2pt, font=\scriptsize, line width=0.4pt,
                minimum height=4.2mm},
  ctarget/.style={cnode, draw=cafeaccent, line width=0.7pt,
                  fill=cafeaccent!8, font=\scriptsize\bfseries},
  coutcome/.style={cnode, draw=cafetruth, line width=0.6pt,
                   fill=cafetruth!6},
  cfrozen/.style={cnode, fill=cafegrid!45, text=cafemuted},
  cedge/.style={-{Stealth[length=1.5mm,width=1.2mm]}, draw=cafemuted,
                line width=0.5pt},
  cpath/.style={cedge, draw=cafeaccent, line width=0.9pt},
  cgone/.style={-{Stealth[length=1.5mm,width=1.2mm]}, draw=cafefaint,
                line width=0.5pt, dash pattern=on 1.2pt off 1.2pt},
  clabel/.style={font=\scriptsize, text=cafemuted, inner sep=1pt},
  pstep/.style={draw=cafemuted, fill=white, rounded corners=2pt,
                line width=0.4pt, inner sep=3pt, font=\scriptsize,
                align=center, minimum height=8mm},
  pkeystep/.style={draw=cafeaccent, fill=cafeaccent!7, rounded corners=2pt,
                  line width=0.7pt, inner sep=3pt, font=\scriptsize,
                  align=center, minimum height=8mm},
  flow/.style={-{Stealth[length=1.6mm,width=1.3mm]}, draw=cafemuted,
               line width=0.55pt},
}

\newtheorem{definition}{Definition}
\newtheorem{problem}{Problem}

\newcommand{\cut}[1]{}

\newcommand{\paraheading}[1]{{\medskip\noindent\textbf{#1.}}}

\newcommand{\para}[1]{\medskip\noindent {\bf #1} }
\newcommand{\tool}{\textit{CAF\'E}}

\newcommand{\Space}[1]{}

\colorlet{shadecolor}{gray!20}
\definecolor{Gray}{gray}{0.8}

\newtheoremstyle{findingstyle}
  {0pt}   %
  {0pt}   %
  {\itshape}  %
  {0pt}       %
  {\bfseries} %
  {.}         %
  {5pt plus 1pt minus 1pt} %
  {}          %

\theoremstyle{findingstyle}
\newtheorem{findinner}{\textbf{Finding}}

\newenvironment{find}
  {\begin{shaded}\begin{findinner}}
  {\end{findinner}\end{shaded}}

\newcommand{\finding}[1]{
  \begin{find}
    #1
  \end{find}
}

\newcommand{\mb}[1]{\mathbf{#1}}

\newcommand{\Pa}{\mb{Pa}}

\newcommand{\numSeedsTab}{5}            %
\newcommand{\numAuditRows}{240}         %
\newcommand{\numImgN}{2{,}000}          %
\newcommand{\numRankIns}{0.919}
\newcommand{\numRankAlarm}{0.932}
\newcommand{\numMaeIns}{0.0105}
\newcommand{\numMaeAlarm}{0.0058}

\newcommand{\numBackMaeMultIns}{6.4}
\newcommand{\numBackMaeMultAlarm}{11.9}

\newcommand{\numAblMeanRankIns}{0.861}
\newcommand{\numAblMeanRankAlarm}{0.875}
\newcommand{\numAblMeanMaeIns}{0.0233}
\newcommand{\numAblMeanMaeAlarm}{0.0153}
\newcommand{\numAblFreshRankIns}{0.914}
\newcommand{\numAblFreshRankAlarm}{0.913}
\newcommand{\numAblFreshMaeIns}{0.0113}
\newcommand{\numAblFreshMaeAlarm}{0.0062}
\newcommand{\numAblMeanMultIns}{2.2}
\newcommand{\numAblMeanMultAlarm}{2.6}
\newcommand{\numAblMeanRankDrop}{0.06}   %

\newcommand{\numDetectPoolN}{60}        %
\newcommand{\numNullReps}{15}           %

\newcommand{\numRetrainN}{15}
\newcommand{\numCorruptIns}{0.890}
\newcommand{\numCorruptAlarm}{0.910}
\newcommand{\numCorruptMaeIns}{0.0298}
\newcommand{\numCorruptMaeAlarm}{0.0147}

\newcommand{\numCorruptRankFloor}{0.89}  %
\newcommand{\numShuffleRankIns}{0.936}
\newcommand{\numShuffleRankAlarm}{0.951}
\newcommand{\numShuffleMaeIns}{0.0151}
\newcommand{\numShuffleMaeAlarm}{0.0079}
\newcommand{\numChanAcc}{0.963}

\newcommand{\numSubAcc}{0.990}

\newcommand{\numBudgetCap}{25{,}000}

\newcommand{\numBudgetRankIns}{0.914}

\newcommand{\numBudgetRankAlarm}{0.875}

\newcommand{\numBudgetParityQ}{480}
\newcommand{\numBudgetShapCap}{0.60}

\newcommand{\numParityGapCapIns}{0.133}    %
\newcommand{\numParityGapCapAlarm}{0.237}

\newcommand{\numSaveIns}{92\%}

\newcommand{\numSaveAlarm}{81\%}

\newcommand{\numImgRankBias}{0.933}
\newcommand{\numHospNone}{0.2189}
\newcommand{\numHospMeta}{0.0370}
\newcommand{\numHospStyle}{0.0127}
\newcommand{\numHospLeace}{0.0148}
\newcommand{\numHospParityCorr}{0.139}    %
\newcommand{\numHospParityCausal}{0.279}  %
\newcommand{\numHospParityMult}{7.5}      %
\newcommand{\numRunGlobalIns}{0.155}
\newcommand{\numRunFullIns}{0.781}
\newcommand{\numRunGlobalAlarm}{0.120}
\newcommand{\numRunFullAlarm}{0.368}
\newcommand{\numAdultTrain}{24{,}129}

\newcommand{\numAdultEdges}{55}
\newcommand{\numAdultSeeds}{3}
\newcommand{\numAdultGraphs}{63}
\newcommand{\numAdultEdu}{32\%}
\newcommand{\numAdultEduOcc}{27\%}
\newcommand{\numAdultSex}{94\%}
\newcommand{\numAdultForest}{52\%}

\begin{document}

\fancyfoot[L]{\footnotesize Preprint. In submission.}
\fancypagestyle{firstpage}{%
  \fancyhf{}%
  \renewcommand{\headrulewidth}{0pt}%
  \renewcommand{\footrulewidth}{0pt}%
  \fancyfoot[L]{\footnotesize Preprint. In submission.}%
}

\begin{abstract}
Machine learning models are increasingly deployed as software components that must evolve as
requirements change. When specific training records or features must no longer influence a
deployed model, machine unlearning aims to remove that influence without retraining from scratch.
Because unlearning is often approximate, its effectiveness must be tested. Such tests must often
treat the model as a black box, without access to its parameters, training history, or unlearning
procedure. Features pose a further challenge: even after a feature is removed from a model's
inputs, its influence can persist through downstream features. Many existing checks examine only
the feature's direct use and can therefore certify a model that still depends on it.

We frame unlearning testing as specification-based testing and present \tool{}, which, using
only a deployed model's predictions, intervenes on the feature, propagates the change to its
downstream features, and checks whether the predictions still respond. \tool{} measures a target's residual influence through both its direct and indirect
causal paths, and its fine-grained diagnostics show which channels and subgroups still carry it.
On two causal-network benchmarks with four unlearning methods, \tool{} ranks residual influence
with 0.92--0.93 pairwise accuracy, against at most 0.71 for existing checks, which fail in both
directions: they certify models whose influence persists through downstream features and flag
correctly unlearned ones. On real census data, \tool{} likewise exposes influence that survives
retraining yet goes unnoticed by direct-input checks.
\end{abstract}

\maketitle
\thispagestyle{firstpage}

\section{Introduction}
\label{sec:intro}

Machine learning (ML) models are increasingly deployed as long-lived software
components. Like other components, they must evolve when requirements change after
deployment. For ML systems, these changes may concern not only functionality but also
the information a model is permitted to use. Privacy requests may require organizations
to remove particular training records; internal policies may prohibit sensitive
attributes or their proxies; and post-deployment debugging may reveal harmful features,
subpopulations, or correlations. Machine unlearning addresses these changes by removing
the influence of targeted information from a trained model efficiently. 
An unlearning request may target individual records,
features, or subpopulations~\cite{xu2023survey,wang2024comprehensive,qu2024learn}.

Unlearning is a model update, and like any update its effectiveness must be tested. This is especially
important for approximate methods, which may leave residual influence. 
Unlike a conventional functional requirement, however, successful unlearning is
not revealed by any single execution: two models, one that has forgotten the target and
one that still depends on it, may return the same prediction on every ordinary test input.
From a software-testing perspective, this is a \textbf{test-oracle problem}~\cite{barr2015oracle}:
ordinary test inputs cannot establish whether the update has removed the target's
influence. Evidence must instead come from comparing predictions on related inputs that a
successfully unlearned model should treat consistently, a metamorphic relation~\cite{segura2016survey}.
Constructing the right related inputs is the central challenge.

Which inputs count as related, and how the updated model must treat them, depends on what
is being unlearned. 
For a \textit{training record}, the required behavior is that the model act as if
it had never seen that record, and existing tests look for its traces through membership
inference, fingerprints, or reference models~\cite{sommer2022athena,shokri2017membership,goel2023towards}.
For a \textit{feature}, the
required property is causal non-influence: 
changing the feature must not change the prediction, 
whether the change reaches the model directly or through features derived from it.
 We therefore frame unlearning testing around target-specific
oracles~\cite{mazhar2026towards} and instantiate this framing for \textit{features}. Even retraining from scratch without the
feature's column, the reference that record-level unlearning treats as ideal, does not
establish this property: the feature's influence can persist through downstream features,
and direct-input tests then report success.

\begin{example}\label{ex:lender}
A default-risk model uses income, credit score, precise location, and commute distance,
which is computed from location and workplace. A policy change prohibits the use of precise
location, so it is unlearned. Afterward, feature attribution assigns location zero importance,
as it would had the model been retrained without that feature. Now suppose an applicant
moves. Under the policy, the model's prediction should not change when their location
changes, yet it does: the move changes their commute distance, which changes the predicted
risk even though location no longer affects it directly (Location $\rightarrow$ Commute
distance $\rightarrow$ Prediction).

\end{example}

\begin{figure}[t]
    \centering
    \begin{tikzpicture}[x=1cm, y=1cm]

\node[font=\scriptsize\bfseries, text=cafetruth, anchor=west] at (0, 1.55)
  {(a) The practitioner's causal specification};

\node[cnode]    (region)  at (0.75,  0.15) {Region};
\node[cnode]    (income)  at (3.00,  1.00) {Income};
\node[ctarget]  (loc)     at (3.00, -0.65) {Location};
\node[cnode]    (commute) at (5.60, -0.65) {Commute};
\node[cnode]    (score)   at (5.60,  1.00) {Credit score};
\node[coutcome] (pred)    at (8.55,  0.15) {Default $Y$};
\node[cnode]    (work)    at (4.30,  0.15) {Workplace};

\draw[cedge] (region) -- (income);
\draw[cedge] (region) -- (loc);
\draw[cpath] (loc)     -- (commute);
\draw[cedge] (work)    -- (commute);
\draw[cpath] (commute) -- (pred);
\draw[cedge] (income)  -- (pred);
\draw[cedge] (score)   -- (pred);
\draw[cedge] (loc.south) -- ++(0, -0.5) -| (pred.south);
\node[clabel, anchor=north, fill=white, inner sep=1.5pt] at (5.80, -1.06)
  {direct causal path};

\node[clabel, anchor=west, align=left, text width=2.6cm]
  at (9.40, 0.15) {Income shares only a common cause, so it is not counted};

\node[font=\scriptsize\bfseries, text=cafetruth, anchor=west] at (0, -2.05)
  {(b) Direct-input probes versus \tool{}'s total probe};

\node[clabel, anchor=east, align=right, text width=2.5cm] at (2.35, -2.75)
  {SHAP, permutation,\\direct perturbation};
\node[ctarget]  (t1) at (3.00, -2.75) {Location};
\node[cfrozen]  (m1) at (5.10, -2.75) {Commute};
\node[coutcome] (p1) at (7.20, -2.75) {Model prediction};
\draw[cgone] (t1) -- (m1);
\draw[cedge] (m1) -- (p1);
\node[clabel, anchor=north, font=\scriptsize] at (5.10, -3.02) {frozen};
\node[clabel, anchor=west, align=left] at (8.25, -2.75)
  {$\tau = 0$\quad\textcolor{cafecausal}{\ding{55}} influence not detected};

\node[clabel, anchor=east, align=right, text width=2.5cm] at (2.35, -3.75)
  {\tool{}, total probe};
\node[ctarget]                    (t2) at (3.00, -3.75) {Location};
\node[cnode, draw=cafeaccent]     (m2) at (5.10, -3.75) {Commute};
\node[coutcome]                   (p2) at (7.20, -3.75) {Model prediction};
\draw[cpath] (t2) -- (m2);
\draw[cpath] (m2) -- (p2);
\node[clabel, anchor=north, font=\scriptsize, text=cafeaccent]
  at (5.10, -4.02) {recomputed from its changed parent};
\node[clabel, anchor=west, align=left] at (8.25, -3.75)
  {$\tau > 0$\quad\textcolor{cafeaccent}{\ding{51}} influence detected};

\end{tikzpicture}
    \caption{The causal specification describes the real-world outcome; the probes test
    the deployed model's prediction of that outcome. Changing location while freezing all
    other inputs produces no model response, but propagating the intervention through
    commute distance changes the prediction. Income remains fixed because its association
    with location is due only to a common cause.}
    \label{fig:overview}
\end{figure}

Existing testing approaches do not expose this failure reliably. Feature attribution and
perturbation methods, including SHAP, permutation importance, and
saliency~\cite{lundberg2017unified,fisher2019modelswrongusefullearning,simonyan2014deepinsideconvolutionalnetworks},
measure predictive dependence but do not by themselves distinguish direct influence,
influence mediated through causally downstream variables, and association caused by
correlation. Fairness testing~\cite{tramer2017fairtest,aggarwal2019bbft,black2020fliptest}
audits discrimination rather than unlearning. Causal-fairness
notions~\cite{kusner2017counterfactual,nabi2018se} provide relevant effect definitions
but are not designed as post-unlearning tests.

We present \tool{}, a causal black-box testing framework for feature unlearning. \tool{}
detects the \emph{residual influence} of an unlearning target: its causal effect on the
updated model's predictions, whether exerted directly or through causally downstream
variables. Beyond a pass or fail verdict,
\tool{} makes a failure actionable. It decomposes the influence into the causal channels, 
enabling targeted remediation, and resolves
it across subgroups, exposing failures that a single score hides. \tool{} 
queries predictions but never inspects parameters, gradients, 
or the unlearning procedure, so it can run as a compliance check in a
release or continuous integration pipeline.  Its only other input is a causal specification of the
 target's downstream features, which for derived features follows from the feature pipeline 
 and need not be exact: a coarse variable ordering suffices on real data, and uncertain edges may be conservatively included: in our experiments, over-declaration raised alarms rather than hiding failures (Section~\ref{sec:robustness-eval}).

At its core, \tool{} is a causal test oracle that turns a causal specification into
metamorphic test inputs. For each observed test instance, it (i)~intervenes on the target,
(ii)~propagates the change through the target's declared descendants, and (iii)~observes the
model's response (Figure~\ref{fig:overview}). A model free of the target's influence must treat the original and
intervened inputs alike. Freezing selected descendants splits the prediction change into
direct and per-channel influence, and repeating the comparison within subgroups shows whom it
affects. Each resulting score is interpretable, tied to a named channel or subgroup, and
remains valid when the model is accessible only as a black box.

We evaluate \tool{} with four unlearning methods on two standard causal-network benchmarks,
externally released reference pairs, a real-world dataset (Adult), and a case
study on real chest X-rays. On the benchmark networks, \tool{} ranks residual influence with 0.92--0.93 pairwise accuracy, against at most 0.71 for existing checks, and stays accurate (0.86--0.89) even when unlearning shrinks influence 20--55$\times$. 
Existing checks fail in both directions: attribution and perturbation
tests certify as fully unlearned every model whose influence persists only through
downstream features, while association- and attack-based checks raise alarms on correctly
unlearned models. Causal knowledge alone does not close this gap: given the same graph,
backdoor adjustment, a standard adjustment strategy for identifying population-level causal effects from observational data,
detects at most half as many failures. On the real-world dataset, given only a coarse
ordering of the variables, \tool{} is the only method that flags every candidate while reporting zero for every clean control.

Beyond detection, \tool{} localizes failures that a single influence score hides, such as
channels whose opposing effects nearly cancel and unlearning that moves influence onto a
previously unaffected subgroup instead of removing it. We further stress-test \tool{}
against errors in the causal graph, the learned mechanisms, and the propagation step, and
find that over-declared edges fail safe, raising alarms rather than hiding failures. Finally,
\tool{}'s query cost is predictable and adjustable: detection alone uses
\numSaveAlarm{}--\numSaveIns{} fewer queries than a full audit with diagnostics, and covering
\tool{}'s \numAuditRows{} test instances costs SHAP about 2.5M queries against
6{,}240--31{,}200 for \tool{}.

\vspace{2mm}
\textbf{Contributions.} This paper makes the following contributions:
\begin{itemize}
    \item We formalize a specification-based framework for unlearning testing built around
    target-specific test oracles: what is being unlearned determines the behavior the updated
    model must satisfy. We instantiate it for feature-level requests using causal
    non-influence within a declared tolerance.

    \item We present \tool{}, a causal black-box testing method for feature unlearning.
    \tool{} operationalizes the feature-level test oracle using paired interventions,
    identifying residual influence and localizing the channels and subgroups through which
    it persists.

    \item We demonstrate \tool{} on standard causal benchmarks, real-world data, and real
    images, and find that it estimates residual influence accurately and exposes unlearning failures that existing checks miss. 
    It also localizes failures hidden by a single score, at low cost.
\end{itemize}

\section{Background and Causal Preliminaries}
\label{sec:background}

We review the forms of machine unlearning and the causal concepts needed to
understand \tool{}. Approaches for testing unlearning are compared in
Section~\ref{sec:related}.

\subsection{Machine Unlearning}
\label{sec:unlearning-definitions}

Machine unlearning updates a trained model so that designated training information no
longer influences it, while avoiding retraining from scratch. Requests may target
individual records, features, or groups of records
\cite{xu2023survey,wang2024comprehensive,qu2024learn}. The target type determines the
behavioral requirement that successful unlearning must satisfy. Methods also differ in
the assurance they provide: exact methods restructure training or maintain sufficient
state so that designated information can be removed without retraining the full model
\cite{cao2015towards,bourtoule2021machine}; certified methods bound how distinguishable
the updated model is from one trained without the designated data
\cite{ginart2019making,guo2020certified}; and approximate methods trade such guarantees
for lower update cost
\cite{graves2021amnesiac,golatkar2020eternal,tarun2023fast}, including in federated
settings \cite{liu2024survey}. Because approximate methods can leave residual influence,
the resulting model must be tested empirically. This paper tests the feature-level
behavior of the updated model, independent of the unlearning algorithm that produced it
(Section~\ref{sec:definitions}).

\subsection{Causal Preliminaries}

\subsubsection{Structural Causal Models}

Let $F$ be the set of observed input features and let $Y$ denote the real-world outcome
that the model predicts. The model prediction $M(x)$ is separate from $Y$. A structural
causal model consists of observed variables $V=F\cup\{Y\}$, exogenous variables $U$, a
distribution $P(U)$, and a structural mechanism $H_X$ for each $X\in V$. Each mechanism
determines $X$ from its observed parents $\Pa_V(X)$ and exogenous parents $\Pa_U(X)$.
The associated causal graph $G=(V,E)$ is directed and acyclic; an edge
$A\rightarrow B$ states that $A$ is a direct cause of $B$
\cite{pearl2009causality}. A descendant of $Z$ is any variable reachable from $Z$ by a
directed path.

In this paper, $G$ is a causal specification supplied by the practitioner (Definition~\ref{def:causal-spec}). \tool{} learns only the mechanisms needed
to propagate interventions through the modeled descendants of the unlearning target.

\subsubsection{Interventions and Instance-Level Counterfactuals}

An intervention $Z\leftarrow z$ replaces the structural mechanism for $Z$ with the
constant $z$ while leaving the remaining mechanisms unchanged. At the population level,
an intervention describes how outcomes change when $Z$ is externally set. \emph{Backdoor
adjustment}, a standard adjustment strategy for identifying such effects from observational data, computes them by averaging over a set of variables that blocks every confounding path
between $Z$ and the outcome~\cite{pearl2009causality}. \tool{} instead
begins with an observed test instance and asks what the model would predict for that same
instance under an alternative value of $Z$, an instance-level counterfactual.

Counterfactual reasoning proceeds through abduction, action, and prediction: infer what
is specific to the observed instance, apply the intervention, and recompute its
descendants \cite{pearl2009causality}. Under an additive-noise mechanism
$X=H_X(\Pa_V(X))+U_X$, the observed residual $U_X$ captures instance-specific information
that can be reused after intervening on the parents. \tool{} uses this principle to
construct approximate instance-level counterfactuals; Section~\ref{sec:algorithm}
describes the propagation procedure and its approximation assumptions.

\subsubsection{Direct, Total, and Channel Effects}

Suppose the target feature $Z$ affects the prediction directly and through mediator
features. A \emph{total effect} allows all descendants of $Z$ to respond to the
intervention, capturing both direct and mediated influence. A \emph{direct effect}
(formally, the controlled direct effect) changes $Z$ while holding its mediators fixed.
For diagnosis, we use \emph{channel influence}, a diagnostic quantity, obtained by propagating the intervention
only along a selected directed path to localize mediated influence. Association created
only by a shared cause is not part of these causal effects.

These notions provide the quantities used in the feature-unlearning test specification
formalized next.
\section{Problem Definition}
\label{sec:definitions}

We treat machine-unlearning testing as specification-based testing.

\begin{definition}[Machine-Unlearning Test Specification]
A \emph{machine-unlearning test specification} identifies (i) the information targeted
for unlearning and (ii) the test oracle that states the behavioral property and
acceptance condition the updated model must satisfy. The target type determines the form
of the test oracle; the test specification instantiates it.
\end{definition}

Record-level targets already have dedicated tests (Section~\ref{sec:related}); feature-level
targets do not. We instantiate the specification for feature-level requests, for which the
required behavior is causal non-influence.

\begin{definition}[Feature Unlearning]
\label{def:feature-unlearning}
Let $M$ be a trained model. A \emph{feature unlearning procedure} is a function
\[
U\colon (M,Z)\longmapsto M'
\]
that takes $M$ and a target feature $Z$ and outputs an updated model $M'$. We say $U$
achieves \emph{effective feature unlearning} if $Z$ has no causal influence on the
predictions of $M'$, either directly or through causally downstream input features.
\end{definition}

This criterion is distinct from retraining equivalence, which is commonly used for
record-level unlearning. Retraining without $Z$'s column does not guarantee causal
non-influence: $Z$ may continue to affect predictions through causally downstream
inputs. Feature-unlearning specifications that require removal of such
descendant-mediated influence therefore require a stronger behavioral condition.
The criterion is behavioral: the updated model may retain information or
representations associated with $Z$, provided that they do not causally affect its
predictions under the declared intervention. We next formalize the evaluation
population, causal specification, intervention contrast, and residual influence used
to instantiate this requirement.

\noindent\textbf{Notation.}
Let $F$ be the set of input features and $Y$ the real-world outcome that the model
predicts. An instance $x$ assigns a value $x[A]\in\textsc{Dom}(A)$ to every
$A\in F$, and a deployed model $M$ maps $x$ to a prediction $M(x)$ of $Y$.
Thus, $Y$ belongs to the causal system, whereas $M(x)$ is the output being tested.

\begin{definition}[Feature Unlearning Target]
\label{def:unlearning-target}
A \emph{feature unlearning target} is a pair $(T,Z)$, where $Z\in F$ is the
\emph{target feature} and $T$ is the finite evaluation population over which the
effectiveness of unlearning is evaluated. The request is global when
$T$ is the full evaluation population and subgroup-specific when $T$ is selected
by a predeclared subgroup.
\end{definition}

We evaluate effectiveness relative to a practitioner-declared causal
specification.

\begin{definition}[Causal Specification]
\label{def:causal-spec}
A \emph{causal specification} is a directed acyclic graph $G=(V,E)$ over
$V=F\cup\{Y\}$. An edge $A\rightarrow B$ allows $A$ to be a direct cause of $B$; a missing edge asserts that it is not.
\end{definition}

The graph determines which features are propagated under an intervention, while
mechanisms determine their intervened values. Let $H$ denote mechanisms compatible
with $G$; in practice, \tool{} estimates them from observational data as $\hat H$.
For an instance $x$ and target value $\theta$, we use the following input notation:
\begin{itemize}
    \item $x^{\mathrm{tot}}_\theta$ sets $Z=\theta$ and recomputes all descendant
    features of $Z$ using the mechanisms;
    \item $x^{\mathrm{dir}}_\theta$ sets $Z=\theta$ and freezes every other feature;
    \item $x^\pi_\theta$ sets $Z=\theta$ and recomputes only the mediator features
    on channel $\pi$.
\end{itemize}

Under the declared specification, variables connected to $Z$ only through shared causes are not propagated, so such associations are excluded from the constructed contrast. Two target values
$(\theta_0,\theta_1)$ form the \emph{contrast} used by a probe. Each probe compares
an intervention pair by querying the model on both inputs. The total pair approximates
the same instance under two alternative interventions; the direct and channel pairs
deliberately restrict propagation for diagnosis.

\begin{definition}[Influence]
\label{def:causal-influence-1}
Let $M$ be a model and $(T,Z)$ a feature unlearning target. For $x\in T$, the
per-instance \emph{total} and \emph{direct influence} of $Z$ are
\[
\tau_{\mathrm{total}}(x)
=M(x^{\mathrm{tot}}_{\theta_1})-M(x^{\mathrm{tot}}_{\theta_0}),
\qquad
\tau_{\mathrm{direct}}(x)
=M(x^{\mathrm{dir}}_{\theta_1})-M(x^{\mathrm{dir}}_{\theta_0}).
\]

For diagnostic localization along a mediated channel $\pi$ from $Z$ through
descendant input features, the \emph{channel influence} is
\[
\tau_\pi(x)
=\big(M(x^\pi_{\theta_1})-M(x^\pi_{\theta_0})\big)
-\tau_{\mathrm{direct}}(x).
\]

The \emph{influence} of $Z$ on $M$ is the mean absolute total influence over $T$:
\[
I(M)
=\frac{1}{|T|}\sum_{x\in T}|\tau_{\mathrm{total}}(x)|.
\]

The diagnostic quantities $I_{\mathrm{direct}}(M)$ and $I_\pi(M)$ are defined
analogously from $\tau_{\mathrm{direct}}$ and $\tau_\pi$. For the post-unlearning
model $M'$, we call $I(M')$ the \emph{residual influence}.
\end{definition}

\tool{} estimates these quantities on the test instances $R\subseteq T$ by replacing
the average over $T$ with an average over $R$.

The mean absolute value prevents opposing effects from canceling across instances.
Channel estimates can likewise reveal mediated effects that offset one another.

For the post-unlearning model $M'$, effective feature unlearning
(Definition~\ref{def:feature-unlearning}) is therefore tested by the condition
\[
I(M')\leq\delta,
\]
where $\delta\geq0$ is the declared tolerance. The tolerance is specified by the practitioner
or governing policy, not by \tool{}, which supplies the estimate of $I(M')$ used in the
comparison. Setting $\delta=0$ is the strict specification for the given population, contrast, and causal specification, which \tool{} estimates on $R$; $\delta>0$ permits an explicitly
tolerated remainder. Direct and channel influence are diagnostic quantities and do not
define separate test outcomes.

\begin{problem}[Feature Unlearning Testing]
\label{prob:unlearning-testing}
Let $M'$ be the result of applying an unlearning procedure to model $M$ for target
$(T,Z)$. Given black-box
access to $M'$, test instances $R\subseteq T$, a causal specification $G$, and
observational data for learning the propagation mechanisms, specify a contrast
$(\theta_0,\theta_1)$ to estimate the residual influence of $Z$ in
$M'$ and determine whether
unlearning is effective in the sense of Definition~\ref{def:feature-unlearning}. When influence is detected, optionally localize the channels and
predeclared subgroups in which it persists. The audit may be subject to a query budget.
\end{problem}

The base audit requires no access to the pre-unlearning model $M$; such access is needed
only for a relative acceptance criterion based on reduction from the original model.
We assume no access to model parameters, gradients, training history, or the internal
operation of the unlearning procedure. The next section presents \tool{}, our solution
to Problem~\ref{prob:unlearning-testing}.
\section{\tool{}: \textit{CA}usal Testing for \textit{FE}ature Unlearning}
\label{sec:framework}

\tool{} operationalizes the feature-level test oracle of
Definition~\ref{def:feature-unlearning}: causal non-influence within the declared
tolerance. It is designed to expose residual influence whether it reaches the prediction
directly or through modeled descendants of the target, using only prediction access at
a predictable and controllable query cost. \tool{} separates detection from diagnosis:
the total probe supplies the audited residual influence, while direct, channel, and
subgroup analyses explain where detected influence persists.

\subsection{Paired Interventions}
\label{sec:algorithm}

\tool{} receives black-box access to the post-unlearning model $M'$, a feature-level
target $(T,Z)$, test instances $R\subseteq T$, and the causal specification $G$; it
chooses the contrast $(\theta_0,\theta_1)$ itself unless one is specified. Each instance in $R$ is a
starting point for paired intervened inputs. The causal specification determines which
features are propagated under an intervention on $Z$. For each modeled descendant feature
$X$, \tool{} fits a mechanism $\hat H_X$ from observational data $D$, so propagation needs only
the target's descendant inputs and their parents. It applies up to three probes to each instance
(Algorithm~\ref{alg:tool}):

\begin{itemize}[leftmargin=*]
    \item \textbf{Total probe.} $Z$ is set to $\theta_0$ and $\theta_1$, and each
    intervention is propagated through all modeled descendant features of $Z$ in
    topological order. The resulting prediction difference estimates per-instance total
    influence.

    \item \textbf{Direct probe.} $Z$ is set to each value while all other variables
    remain fixed at their observed values. The resulting prediction difference estimates
    per-instance direct influence.

    \item \textbf{Channel probe.} For a channel $\pi$, a selected directed path from
    $Z$ through descendant input features, the intervention is propagated only through
    features on that path. Subtracting direct influence from the resulting prediction
    difference estimates that channel's mediated contribution.
\end{itemize}
Each probe produces the per-instance quantity defined in Section~\ref{sec:definitions}.
The total probe supplies $I(M')$; the direct and channel probes show whether detected
influence remains at the target input or travels through particular descendants. Channels
may share features, so their estimates localize influence rather than partition the total.
Because every influence is computed per instance, subgroup analysis needs no further queries:
for a predeclared subgroup $S\subseteq T$, \tool{} averages the same per-instance estimates over
$R\cap S$.

\newcommand{\INPUT}{\item[\textbf{Input:}]}
\newcommand{\OUTPUT}{\item[\textbf{Output:}]}
\begin{algorithm}[t]
\caption{\tool{}: paired-intervention testing}
\label{alg:tool}
\begin{algorithmic}[1]
\INPUT
Causal graph $G$, black-box model $M'$, target $Z$, test instances $R$,
observational data $D$
\OUTPUT Per-instance $\tau_{\text{total}}$, $\tau_{\text{direct}}$, and $\tau_{\pi}$ for each channel $\pi$
\STATE $(\theta_0,\theta_1) \gets$ contrast from $D$
\STATE $\hat H \gets$ fit $\hat H_X$ from $D$ for each $X\in\textsc{Desc}_F(Z)$
\FOR{each instance $x \in R$}
    \FOR{$\theta \in \{\theta_0,\theta_1\}$}
        \STATE $x^{\text{tot}}_{\theta} \gets \textsc{Propagate}(x, Z{\leftarrow}\theta, \textsc{Desc}_F(Z), \hat H)$
        \STATE $x^{\text{dir}}_{\theta} \gets x$ with $Z{\leftarrow}\theta$ \COMMENT{all else frozen}
    \ENDFOR
    \STATE $\tau_{\text{total}}(x) \gets M'(x^{\text{tot}}_{\theta_1}) - M'(x^{\text{tot}}_{\theta_0})$
    \STATE $\tau_{\text{direct}}(x) \gets M'(x^{\text{dir}}_{\theta_1}) - M'(x^{\text{dir}}_{\theta_0})$
    \FOR{each selected channel $\pi$ from $Z$}
        \STATE $x^{\pi}_{\theta} \gets \textsc{Propagate}(x, Z{\leftarrow}\theta, \textsc{Nodes}_F(\pi){\setminus}\{Z\}, \hat H)$ for $\theta\in\{\theta_0,\theta_1\}$
        \STATE $\tau_{\pi}(x) \gets \big(M'(x^{\pi}_{\theta_1}) - M'(x^{\pi}_{\theta_0})\big) - \tau_{\text{direct}}(x)$
    \ENDFOR
\ENDFOR
\RETURN per-instance influences and their summaries
\end{algorithmic}
\end{algorithm}

\paraheading{Propagation}
\textsc{Propagate} updates the selected descendant features in topological order and keeps
each instance's own unexplained variation, so each intervened input approximates that instance's counterfactual by standard abduction, action, and prediction~\cite{pearl2009causality}; \tool{} operationalizes it as a black-box test oracle. For a continuous additive-noise
mechanism, \tool{} reuses the observed instance's residual (the part of its value
unexplained by its parents) under the intervened parents. For a categorical descendant,
whose observed value does not identify a unique latent draw, \tool{} averages the model's
predictions over compatible latent draws (two per intervened input in our experiments;
Algorithm~\ref{alg:tool} omits these repeated draws). Errors in the fitted mechanisms change
only the size of the estimated influence, never whether a model that ignores the target is
flagged (Section~\ref{sec:decision}).

\paraheading{Contrasts}
Each contrast $(\theta_0,\theta_1)$ is a test case: a nonzero residual influence at any single
contrast is a counterexample to effective unlearning. By default, \tool{} uses the target's two
values for a binary target and its 10th and 90th percentiles for a continuous one. Because the
probe works with any contrast, a practitioner can add contrasts, or generate them adaptively from the
model's responses, to search for larger residual influence, at a query cost linear in the number
of contrasts (Section~\ref{sec:cost}).

\subsection{From Residual Influence to an Audit Decision}
\label{sec:decision}

\tool{} estimates $I(M')$ as the mean absolute per-instance total influence over $R$ and
compares it with the declared tolerance $\delta$ of Definition~\ref{def:feature-unlearning}.
Our experiments use the strictest tolerance, $\delta=0$: \tool{} bootstrap-resamples the test
instances 1{,}000 times and detects residual influence when the one-sided lower bound
(5th percentile) on mean absolute influence exceeds zero. This rule never flags a deterministic model that ignores the target and its declared
descendants, however accurate the fitted mechanisms. Every constructed pair then receives
the same prediction, so the estimate and every resample are exactly zero. Each detection is therefore evidence of a violation under the declared specification.

\subsection{Query Cost}
\label{sec:cost}

A detection-only audit requires one total pair per test instance. \tool{}'s default base
audit additionally uses a direct pair for coarse diagnosis, and each selected channel
requires one further pair. Query cost is therefore linear in the number of test instances
and diagnosed channels; measured counts include the repeated draws for categorical
descendants.

A practitioner can therefore stage the audit: first test for residual influence, then use the
direct probe and selected channel probes when diagnosis is needed. Under $\delta=0$ any
residual influence triggers diagnosis; under $\delta>0$, offsetting channels can keep the total
below the tolerance, so a practitioner can also run channel probes on models that pass narrowly
(Section~\ref{sec:localization-eval}), balancing diagnostic detail and 
query cost as quantified in Section~\ref{sec:cost-eval}.
\section{Evaluation}
\label{sec:evaluation}

We evaluate \tool{} on two benchmark causal networks, real tabular data, and X-ray images,
against seven baselines. Throughout, \tool{} has only black-box access to each audited model,
together with the causal specification and the data used to fit its propagation mechanisms.

    \para{RQ1 (Detecting Residual Influence):}
    How accurately does \tool{} detect, estimate, and rank residual influence relative to
    direct-input, causal, and association- and attack-based baselines?

    \para{RQ2 (Localizing Failures):}
    When residual influence is present, do channel diagnostics and subgroup analyses
    identify where it persists?

    \para{RQ3 (Robustness):}
    How does \tool{} degrade under errors in the graph, the fitted mechanisms, or propagation?

    \para{RQ4 (Cost):}
    How many model queries does an audit need, and how does accuracy
    change with the query budget?

\subsection{Experimental Setup}

\paraheading{Datasets and ground truth}
Each testbed has its own ground truth, and we report results per testbed rather than pooled.
(1)~\emph{Insurance}~\cite{binder1997adaptive} (27 nodes, 52 edges) and
\emph{ALARM}~\cite{alarm} (37 nodes, 46 edges), standard Bayesian networks from the bnlearn
repository~\cite{scutari2010learning}, test the full audit: detection, ranking, effect magnitude,
and localization when channels overlap. \tool{} receives the published graph but never the true
mechanisms; for ground truth, we regenerate each test instance under both values of the default
contrast from those mechanisms, keeping its noise fixed, and query the same candidate model on the
pair.
(2)~\emph{Hospital X-ray} tests whether \tool{} separates causal from merely correlated influence
when the target is a concept: hospital identity, in a model predicting infiltration from
\numImgN{} balanced ChestMNIST~\cite{yang2023medmnist} $64{\times}64$ images. A controlled scanner
mechanism renders the same anatomy in either hospital's style, giving same-anatomy reference pairs.
(3)~\emph{Adult}~\cite{adult_2} tests detection on real data, where no correct graph is known:
\tool{} receives only an ordering of the variables, and clean controls, whose influence is zero under that ordering, serve as ground truth (RQ1
and RQ3).
(4)~\emph{Controlled SCMs} plant failures whose channels and subgroups are known by construction,
for localization (RQ2) and mechanism misspecification (RQ3).

\paraheading{Models and unlearning}
Audited models are a two-layer MLP (64 and 32 units) on tabular data and a two-layer CNN (16 and
32 channels) on images. For each Insurance and ALARM seed, we generate 1{,}200 instances, with a
stratified split of 960 for fitting the model and \tool{}'s mechanisms (logistic regressions on
these categorical networks, linear regressions for continuous descendants elsewhere) and
\numAuditRows{} for testing. Each network has one unlearning target, fixed in advance, independently of any measured effect. On
Insurance and ALARM we apply four unlearning methods: retraining from scratch without the target's
column, feature shuffling~\cite{yang2025featureunlearning} with fine-tuning,
FOUL~\cite{warnecke2023unlearning}, and MIUL~\cite{guo2022efficient}; on Adult, retraining without
the target's column; on images, metadata removal, paired-style augmentation, and
LEACE~\cite{belrose2023leace}.
The \emph{original model} is the one before unlearning; a \emph{candidate model} is what an
unlearning method returns, and is what \tool{} audits; a \emph{clean control} scrubs the target and
all its descendants, so its residual influence is exactly zero by construction. We do not select
candidates by effect size: the pool contains whatever effects the methods leave behind.

\paraheading{Baselines and metrics}
The baselines fall into three classes, each testing a different reason an existing check
might already suffice (Section~\ref{sec:related}).
\emph{Direct-input probes}---SHAP~\cite{lundberg2017unified}, permutation
importance~\cite{fisher2019modelswrongusefullearning}, conditional
permutation~\cite{strobl2008conditional}, and direct perturbation---measure the model's response to
the target's own input, as attribution and attribute-flip fairness tests do; direct perturbation is
exactly \tool{} without propagation (Section~\ref{sec:algorithm}).
\emph{Backdoor adjustment}~\cite{pearl2009causality} (Section~\ref{sec:background}), uses \tool{}'s graph to identify the same contrast; our g-computation estimate is population-average rather than paired and instance-level.
\emph{Association-based checks}, from the fairness and unlearning-verification literatures, are
statistical parity difference~\cite{bellamy2018aifairness360extensible} (the largest gap in
positive-prediction rate between the target's groups) and the attribute inference attack (an
attacker trained on the model's output probabilities); they never perturb the input, so they still
apply after the target's column is removed.

Every method receives the same 500{,}000-query ceiling, excluding queries used to build the ground
truth, and queries the model's positive-class probability, as \tool{} does. In the detection study,
KernelSHAP (256 samples per instance) instead receives 3M queries so that it sees the same
\numAuditRows{} test instances as every other method. To score estimates against the ground truth, each
method estimates the residual influence of every feature in the network, not only the target. Most
comparators report scores on their own scales, so we compare them by \emph{pairwise ranking
accuracy}: the fraction of feature pairs with different reference effects that a method orders
correctly. For the methods that estimate the reference
effect itself, we also report effect mean absolute error (MAE). Detection is scored by detection
and false-alarm rates.
Ranking, robustness, and cost results average \numSeedsTab{} seeds, as does the image study;
detection uses 15 seeds per network.

\subsection{Detecting Residual Influence (RQ1)}
\label{sec:detection-eval}

\tool{} must deliver three things. First, a verdict: flag every model that still uses the target and pass every model that does not. Second, correct influence estimates for the other features as well, since the target's indirect influence reaches the model through them; we score these estimates here, and Section~\ref{sec:localization-eval} uses them to localize the channels that carry it. Third,
both of these when the target is a concept rather than a column. \tool{} delivers all three, on benchmark networks, real tabular data, and real X-ray images. Each family of comparators falls short for its own reason, and the same reason recurs
across all three.

\paraheading{Catching every failure without a false alarm}
We audit \numDetectPoolN{} candidates per network from the four unlearning methods, each with
nonzero residual influence, and \numNullReps{} clean controls, which have exactly zero, under
strict detection ($\delta=0$). \tool{} detects every candidate on both networks with no false alarm
(Table~\ref{tab:detection}), including FOUL's, whose residuals are the smallest in the pool, down to
$10^{-5}$. Its zero false alarms are structural: a clean control reads neither the target nor its
descendants, so under the correct graph and a deterministic model \tool{}'s estimate is exactly
zero; RQ3 relaxes both conditions. The direct-input probes fail on the most common unlearning
method. Retraining without the target's column removes its direct influence entirely, yet 84\%
(Insurance) and 93\% (ALARM) of its influence survives through its descendants. The probes measure
only the direct part, so they report exactly zero on every retrained model and would certify each
one as forgotten. Direct perturbation is \tool{} without propagation (Section~\ref{sec:algorithm}), so this gap is propagation's value.

\begin{table}[t]
    \centering
    \caption{Strict detection ($\delta=0$) rates, as Insurance\,/\,ALARM. Per network:
    \numDetectPoolN{} candidates, \numRetrainN{} of them retrained without the target's column, and
    \numNullReps{} clean controls. \tool{} and the direct-input probes fire when a bootstrap lower bound on
    mean absolute influence exceeds zero; the statistical baselines, when their statistic exceeds 95\% of a
    permutation null. Detection asks only whether influence is nonzero; ranking (Figure~\ref{fig:ranking-by-treatment}) is the harder test.}
    \label{tab:detection}
    \begin{tabular}{lccc}
        \toprule
        & \makecell{\textbf{Retrained without}\\\textbf{target column} $\uparrow$}
        & \makecell{\textbf{All}\\\textbf{candidates} $\uparrow$}
        & \makecell{\textbf{False alarms on}\\\textbf{clean controls} $\downarrow$} \\
        \midrule
        Direct-input probes$^{*}$ & 0.00\,/\,0.00 & 0.75\,/\,0.75 & 0.00\,/\,0.00 \\
        Backdoor adjustment & 0.47\,/\,0.00 & 0.50\,/\,0.02 & 0.00\,/\,0.07 \\
        Statistical parity difference & 0.27\,/\,0.00 & 0.18\,/\,0.00 & 0.40\,/\,0.00 \\
        Attribute inference attack & 0.13\,/\,0.07 & 0.27\,/\,0.12 & 0.33\,/\,0.13 \\
        \textbf{\tool{}} & \textbf{1.00\,/\,1.00} & \textbf{1.00\,/\,1.00} & \textbf{0.00\,/\,0.00} \\
        \bottomrule
    \end{tabular}

    \vspace{2pt}
    {\footnotesize $^{*}$SHAP, permutation importance, conditional permutation, and direct
    perturbation; all four give the same detection result.}
\end{table}

The statistical baselines need no column to perturb, but they infer influence from how predictions
vary with the target's observed values rather than intervening on it, and this fails in both
directions. Statistical parity difference and the attribute inference attack measure association,
not influence: their detection rates are at most 0.27 on either network, yet their false-alarm rates on Insurance clean controls are 0.40 and 0.33, where the target has causes the clean model still reads and so remains
associated with, and recoverable from, predictions it does not influence. A correctly unlearned
model can therefore fail them. Backdoor adjustment receives the same graph and targets the same contrast as \tool{}, yet its detection rates are only 0.50 on Insurance and 0.02 on ALARM, so the graph alone is not enough. All three depend on how the target's values
happen to be distributed: on ALARM the target takes two of its three values in only 15 of
\numAuditRows{} test instances, so they detect almost nothing there and raise fewer false alarms
than on Insurance. \tool{} and the direct-input probes set the target's value by intervention
instead.

\begin{figure}[t]
    \centering
    \includegraphics[width=\linewidth]{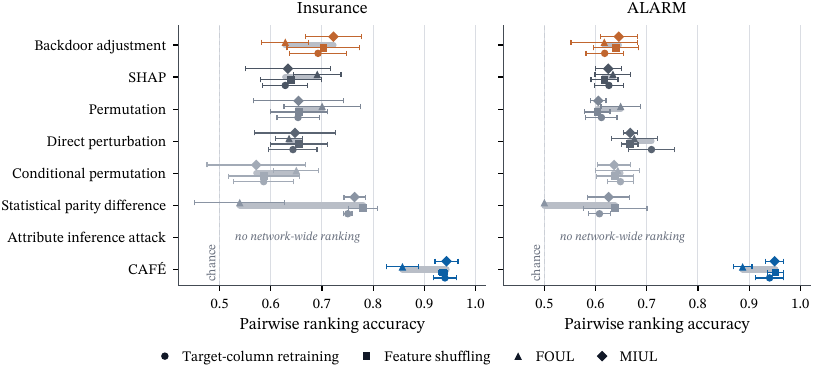}
    \caption{Pairwise ranking accuracy of every feature's estimated residual influence against its
    reference effect (0.5 is chance), after each unlearning method, on Insurance and ALARM.
    One mark per unlearning method, with the bar showing $\pm$1 SD over \numSeedsTab{} seeds and the
    grey spine the spread across methods. The attribute inference attack has no network-wide
    ranking: rare feature values leave its attacker too few examples to train on. \tool{} ranks
    most accurately after every unlearning method, on both networks.}
    \label{fig:ranking-by-treatment}
\end{figure}

\paraheading{On real data}
The same failures appear on the real Adult dataset~\cite{adult_2}. We train a model to predict
income from \numAdultTrain{} records and unlearn a feature by retraining without its column, in
four cases: education; education together with its strongest proxy, occupation; sex; and education
from a random forest instead of an MLP. Each target also gets a clean control, retrained without
the target and every later variable, which its influence cannot reach under the declared ordering. \tool{}'s only causal input
is a five-tier ordering of the variables (demographics, education, family status, employment, then
hours and capital), expanded into a complete graph. The direct-input probes report exactly zero
for every candidate and would certify each one as forgotten. \tool{} estimates that \numAdultEdu{} of education's pre-unlearning influence remains after retraining, \numAdultEduOcc{} when occupation is removed alongside it, and \numAdultForest{} in the random forest. The starkest case is sex: \tool{} estimates that \numAdultSex{} of its influence remains, as propagation passes through the \texttt{relationship} column, which records husband or wife and is highly predictive of sex. \tool{} flags every candidate and reports exactly zero for both controls,
while the statistical baselines again fail in both directions: backdoor adjustment flags 11 of 12
candidate audits but also the sex control in every seed, and the attribute inference attack flags
only 7 candidates yet fires on the controls in five of six audits. Statistical parity raises no false alarm but misses a candidate. Only \tool{} flags every candidate and no control.

\paraheading{Estimating residual influence accurately}
An estimator that misjudges the features carrying the target's influence cannot be trusted on the
target itself. Scoring every feature of each network against its reference effect,
Figure~\ref{fig:ranking-by-treatment} reports each method's pairwise ranking accuracy after each
unlearning method. \tool{} reaches
0.94--0.95 after retraining, feature shuffling, and MIUL on both networks, with no seed below 0.90;
no comparator averages above 0.78, and the comparators' worst seeds fall to chance (0.47--0.51).
The detection failures carry over: the direct-input probes rank at 0.57--0.71 after every unlearning method, since about half of all influence in these networks flows through descendants. Backdoor adjustment ranks at 0.62--0.72 and
has \numBackMaeMultIns{}--\numBackMaeMultAlarm{}$\times$ \tool{}'s effect MAE (\numMaeIns{} and
\numMaeAlarm{}). \tool{}'s magnitudes also hold against reference pairs released independently of
this work: on 1{,}000 IncomeSCM~\cite{johansson2024incomescm} pairs, it matches the reference effect
to within $5\times10^{-10}$ before and after unlearning, while backdoor adjustment is off by 0.0081
and 0.0065, as large as the effects themselves (0.0106 and 0.0056).

FOUL is the hardest case to rank: it shrinks every feature's influence 20--55$\times$, leaving most
features within 0.005 of one another. \tool{} still ranks them at 0.86 and 0.89 (\numRankIns{} and
\numRankAlarm{} averaged over all four methods), but statistical parity difference, the strongest
comparator on Insurance after the other three methods (0.75--0.78), falls to chance (0.54 and
0.50). Parity compares positive-prediction rates across a feature's groups, so it registers
influence only when that influence flips a predicted label. FOUL's effects rarely flip one, so nearly every gap is zero and the features tie.

\paraheading{Separating causal from correlated influence for a concept}
Unlearning requests often target a concept with no column of its own, such as the hospital an
image came from. Here, hospital identity reaches the model through scanner style, and removing the
hospital metadata leaves a scanner-mediated effect of \numHospMeta{} (untreated: \numHospNone{}).
\tool{} reports it, while a metadata-only check, like the direct-input probes, reports zero.
Neither paired-style augmentation (\numHospStyle{}) nor LEACE (\numHospLeace{}), which makes a
concept linearly unrecoverable, removes it: erasing what a model represents does not remove what
drives its predictions. \tool{} fits the scanner mechanism's two parameters (image contrast and
grid amplitude) by least squares from 64 noisy calibration scans; the reference pairs come from a
hidden renderer and are never shown to \tool{}.

In a correlated control, hospital and scanner style share only a common cause, so the true effect
after metadata removal is zero, and \tool{} reports zero. The association-based checks fail here as
they did on the networks. A linear probe on the image embeddings still predicts hospital (0.724
AUROC), so a recoverability check would flag a model that hospital does not drive; the
feature-unlearning criterion is behavioral (Section~\ref{sec:definitions}), and a model may retain
information about the target as long as it does not act on it. Statistical parity difference
reports \numHospParityCorr{} here and \numHospParityCausal{} in the causal condition,
\numHospParityMult{}$\times$ the true \numHospMeta{}, so it cannot tell the two conditions apart.
TCAV~\cite{kim2018tcav} measures whether a concept is represented, not whether it drives
predictions; after metadata removal its hospital sensitivity is indistinguishable from
random-concept controls (empirical $p=0.324$).

\phantomsection\finding{
\tool{} detects every unlearning failure with no false alarm, ranks residual influence at 0.94--0.95 across unlearning methods, and separates causal from correlated influence for a
concept-level target. On real Adult data, where every direct-input check reports zero, \tool{} estimates that \numAdultSex{} of the influence of sex survives retraining under the declared ordering. Each comparator family fails for its own reason: direct-input checks miss
influence that flows through descendants, association-based checks confuse association with
influence in both directions, and backdoor adjustment, given the same graph, lacks the precision of
paired interventions.}

\subsection{Localizing Unlearning Failures (RQ2)}
\label{sec:localization-eval}

Section~\ref{sec:detection-eval} showed that \tool{} estimates residual influence accurately for every feature; RQ2 uses those
estimates to show where the target's residual influence sits. A target usually reaches the model
through several channels, each calling for a different repair, such as retiring a derived feature
or replacing a proxy, and its influence can fall unevenly across people. \tool{} therefore
localizes along two axes: channel diagnostics ask through which channel influence persists, and
subgroup diagnostics ask for whom. We score both against references computed from the true
mechanisms with each instance's noise held fixed, which \tool{} never sees, on candidates from the
four unlearning methods and a clean control over five seeds. A channel or subgroup is active when
its residual influence exceeds both 10\% of its original influence and 0.01; the same rule labels
the reference effects and \tool{}'s estimates.

\begin{figure}[t]
    \centering
    \includegraphics[width=\linewidth]{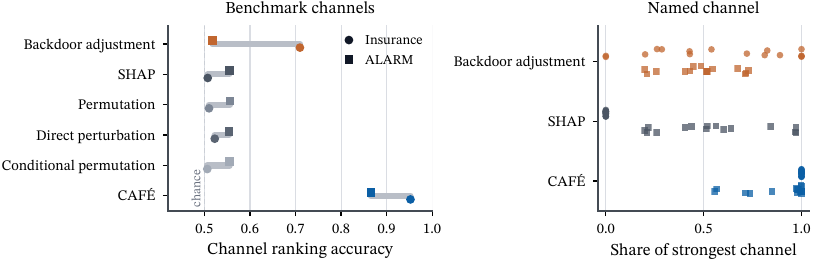}
    \caption{Left: channel ranking accuracy on the benchmark networks, whose channels share
    mediators (0.5 is chance); feature-scoring baselines stay at chance. Right: each mark is one model; it shows how much influence is carried by the channel a method picks as
    strongest, relative to the true strongest channel (1 means the method picked it). On ALARM, where
    channels nearly tie, \tool{}'s picks stay close to 1.}
    \label{fig:localization}
\end{figure}

\paraheading{Channels on the benchmark networks}
The Insurance and ALARM networks from Section~\ref{sec:detection-eval} are the realistic test:
the target reaches the model through 66 and 47 channels that share mediators, so \tool{} must tell
channels apart where they overlap. For each of the 25 audited models per network, we estimate the
residual influence carried by every channel and compare it with a reference that replays the
published mechanisms along that channel alone; the feature-scoring baselines are credited with the
score of each channel's last mediator. \tool{} ranks channels by residual influence with pairwise accuracy 0.952 on Insurance and 0.865 on ALARM (Figure~\ref{fig:localization}, left), with effect MAE at or below 0.003 on both. On Insurance it names the strongest
channel in all 16 models with an active channel; after retraining without the target's column, for
example, it traces the remaining influence along
\texttt{RiskAversion}$\to$\texttt{SeniorTrain}$\to$\texttt{DrivingSkill}$\to$\texttt{DrivQuality}$\to\cdots$ 
matching the reference effects. On ALARM many channels carry nearly equal
influence, so the useful question is whether the named channel is worth repairing: \tool{} names the strongest in 4 of 15 models, and when it misses, its pick carries a median 97\% (at least 56\%) of the strongest channel's residual influence (Figure~\ref{fig:localization}, right). SHAP and the other
baselines score input features rather than channels, so they cannot separate channels that share a
mediator: they rank the benchmark channels at chance (pairwise accuracy 0.51--0.56) and never name
the strongest one.

\paraheading{Seeded failures}
Scoring localization requires knowing where the failure is, and some failures a practitioner must
recognize cannot be produced on demand in a fixed network. We therefore plant failure patterns
with known locations in controlled SCMs, as fault-injection studies do for fault localization:
influence through a single proxy, through one or two of several channels, through channels that
offset each other, or through none; and subgroup effects that are concentrated, opposed, unequal,
or confined to a rare group. \tool{} names the channel carrying the most residual influence in
\numChanAcc{} of cases and the most affected subgroup in \numSubAcc{}, without missing an active
channel or group or flagging an inactive one. It estimates each channel's residual influence to
0.0050 effect MAE and each group's to 0.0026, $14\times$ and $18\times$ closer than backdoor
adjustment.

\paraheading{What a population figure hides}
The seeded failures also show why localization matters: in each of the following cases, a single
population figure misleads. In the cancellation scenario, two channels carry $+0.53$ and $-0.50$
after retraining, each almost untouched by unlearning. Strict detection still flags the model, but
the population figure is only the small difference between the two, so a tolerance above that
difference would pass a model that still relies heavily on the target, and a later update to
either channel would expose it; only the channel view shows how much influence remains. When the
effect sits in a group that is 10\% of the population, retraining leaves a population residual of
0.076, while \tool{} reports 0.39 for that group, five times the average. Under FOUL, the population figure falls by 30\% and the affected group improves, which reads as progress, yet
\tool{} shows that a group that was barely affected (0.07) now carries 0.28, in every seed: the
procedure moved the influence rather than removing it.

\phantomsection\finding{
\tool{} localizes residual influence on real benchmark networks, ranking channels that share
mediators at 0.952 and 0.865 where every baseline is at chance, and on seeded failures it names the
responsible channel and subgroup without a false flag, exposing offsetting channels and
concentrated or relocated influence that a population figure hides.}

\subsection{Robustness (RQ3)}
\label{sec:robustness-eval}

\tool{} relies on three inputs a practitioner supplies or fits: the causal graph, the propagation
mechanisms, and the propagation step that combines them. We corrupt each in turn and ask what
survives: detection, ranking, or effect magnitude. No error we introduced caused a missed failure;
errors cost magnitude first and ranking only slightly.

\begin{table}[t]
    \centering
    \caption{Propagation variants with the graph and fitted mechanisms held fixed: pairwise ranking
    accuracy / effect MAE. All three detect every candidate with no false alarm on the clean controls.}
    \label{tab:propagation-ablation}
    \begin{tabular}{lcc}
        \toprule
        \textbf{Propagation} & \textbf{Insurance} & \textbf{ALARM} \\
        \midrule
        Fresh noise & \numAblFreshRankIns{} / \numAblFreshMaeIns{} & \numAblFreshRankAlarm{} / \numAblFreshMaeAlarm{} \\
        Conditional mean & \numAblMeanRankIns{} / \numAblMeanMaeIns{} & \numAblMeanRankAlarm{} / \numAblMeanMaeAlarm{} \\
        \textbf{Own noise (\tool{})} & \textbf{\numRankIns{} / \numMaeIns{}} & \textbf{\numRankAlarm{} / \numMaeAlarm{}} \\
        \bottomrule
    \end{tabular}
\end{table}

\paraheading{Graph errors}
Under-declaring the graph can hide influence only if it removes every declared route from the
target to the model: deleting the target's edge to its most influential child missed no
candidate, because other routes still carried the influence. Over-declaring is conservative: it
never costs a detection, and its only cost is false alarms. A spurious edge from the target to a feature a clean control
reads makes \tool{} fire on 93\% (Insurance) and 80\% (ALARM) of clean controls, and randomly
adding, deleting, or reversing 20\% of edges on 27\% and 40\%. These errors fail safe, raising an
alarm for review rather than passing a failed model, and the alarm points at its cause: \tool{}'s
channel diagnostics attribute every such false alarm to the wrongly declared feature, and a
likelihood-ratio test of that edge, fitted with \tool{}'s own mechanisms, rejects every spurious
edge while rejecting a true edge in only 3\% of checks on the correct graph. A practitioner unsure of an edge can therefore declare it conservatively. Graph errors also leave the
ranking largely intact (Table~\ref{tab:graph-corruption}): under 5--20\% random corruption it never falls below
\numCorruptRankFloor{}, against \numShuffleRankIns{} and \numShuffleRankAlarm{} with the correct
graph, while effect MAE roughly doubles at 20\%.

The most realistic graph error is a coarse one. On Adult (Section~\ref{sec:detection-eval}),
\tool{} receives only a variable ordering, expanded into a complete graph in which every variable
may affect every later one (\numAdultEdges{} edges); we also delete 10\% or 20\% of its edges at
random, \numAdultGraphs{} graphs in all over \numAdultSeeds{} seeds. Under every graph, \tool{}
detects every candidate and reports exactly zero for both clean controls: deleting a fifth of the
edges changes how much influence it attributes, but never whether it finds it.

\begin{table}[t]
    \centering
    \caption{Rank accuracy / effect MAE under random edge corruption (feature shuffling).
    Matching correct-graph references are \numShuffleRankIns{} / \numShuffleMaeIns{} on Insurance and
    \numShuffleRankAlarm{} / \numShuffleMaeAlarm{} on ALARM.}
    \label{tab:graph-corruption}
    \small
    \begin{tabular}{lccc}
        \toprule
        \textbf{Network} & \textbf{5\%} & \textbf{10\%} & \textbf{20\%} \\
        \midrule
        Insurance & 0.928 / 0.0162 & 0.914 / 0.0248 & \numCorruptIns{} / \numCorruptMaeIns{} \\
        ALARM & 0.944 / 0.0125 & 0.918 / 0.0172 & \numCorruptAlarm{} / \numCorruptMaeAlarm{} \\
        \bottomrule
    \end{tabular}
\end{table}

\paraheading{Mechanism errors}
\tool{} fits a mechanism for each descendant (Section~\ref{sec:algorithm}), and a practitioner
rarely knows the true functional form. In a controlled SCM with nonlinear mediators, a misfit
linear mechanism still orders the audited models correctly in all but one of 30 comparisons,
and a random forest fit recovers the effects closely (effect MAE about 0.02, against 4--5$\times$
that for the linear fit). The image case study shows the same pattern: a
20\% bias in the scanner mechanism raises effect MAE to about a third of the true \numHospMeta{}
effect, while ranking accuracy stays at \numImgRankBias{}. A wrong mechanism thus costs magnitude,
not ordering: a practitioner who needs effect sizes should fit flexible mechanisms, while one who
only needs to compare models can tolerate a crude fit.

\paraheading{Propagation shortcuts}
\tool{} recomputes each descendant from its fitted mechanism and adds back the instance's own noise
term (Section~\ref{sec:algorithm}). Keeping the graph and mechanisms fixed, we either replace that
noise with a random draw (\emph{fresh noise}) or drop it (\emph{conditional mean};
Table~\ref{tab:propagation-ablation}). Every variant still detects all \numDetectPoolN{}
candidates on each network with no false alarm. Ranking falls by about \numAblMeanRankDrop{}
without the noise term and by at most 0.02 with a random one. Effect magnitude depends on it
most: dropping the noise term raises effect MAE \numAblMeanMultIns{}--\numAblMeanMultAlarm{}$\times$,
while a random draw costs under 10\%, since on these categorical networks an observed category
constrains the noise only loosely. The noise term shapes how large an effect
\tool{} reports, not whether it finds one.

\paraheading{Small effects and noisy predictions}
Strict detection is highly sensitive: blending candidates into clean controls, \tool{} detects
every blend, down to influence of $2\times10^{-8}$, which it estimates within 7\% of the truth. It also keeps zero false
alarms on a noisy prediction interface, provided repeated queries are cached: a clean control reads
neither the target nor its descendants, so both inputs of each pair reach it as the same query and
receive the same answer. Without caching, fresh noise on every prediction would register as
influence on the clean controls.

\phantomsection\finding{
No graph, mechanism, or propagation error we introduced cost a detection: errors degrade effect
magnitude first, while ranking stays at or above 0.86. Over-declaring the graph fails safe: it
raises an alarm that \tool{}'s own channel diagnostics trace to the wrong edge, rather than passing
a failed model, and a coarse variable ordering sufficed for detection on Adult.}

\subsection{Cost (RQ4)}
\label{sec:cost-eval}

An auditor often reaches a deployed model only through a prediction API that is rate-limited or
billed per call, so an audit is practical only if its query cost is low and adjustable. \tool{}'s
is both: a full audit costs 80--400$\times$ fewer queries than SHAP needs for the same test
instances and runs in under a second, and a smaller budget costs little accuracy.

\begin{figure}[t]
    \centering
    \includegraphics[width=\linewidth]{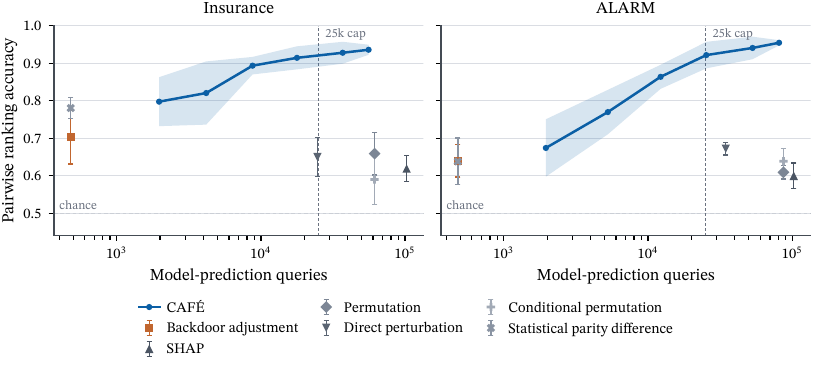}
    \caption{Pairwise ranking accuracy over all features against the queries each audit actually
    used. \tool{}'s curve is the budget sweep, one point per cap, plotted at its measured cost, which
    falls below the cap; each comparator is one mark at the largest predeclared test set it can afford,
    with $\pm$1 SD over \numSeedsTab{} seeds. Statistical parity difference is the cheapest
    method in the comparison at \numBudgetParityQ{} queries, but its accuracy is fixed: buying
    more budget cannot improve a group rate gap that already uses every test instance. The attribute inference attack is omitted: it has no network-wide ranking (Section~\ref{sec:detection-eval}).
    \tool{} ranks most accurately at every budget.}
    \label{fig:accuracy-vs-budget}
\end{figure}

\paraheading{What a full audit costs}
Counting one query as one prediction for one input, auditing one target over all \numAuditRows{}
test instances costs \tool{} 6{,}240--31{,}200 queries. Covering the same instances costs
KernelSHAP, run with enough samples to be stable (256 per instance), about 2.5M; within the
500{,}000-query ceiling every other method receives, it covers only 20 test instances and detects
45\% of ALARM candidates. \tool{}'s counts include queries to the original model, which the
strict detection criterion does not need, and to channel diagnostics, which are optional: omitting
them cuts network-wide query cost by \numSaveIns{} on Insurance and \numSaveAlarm{} on ALARM, so a
practitioner can detect first and localize only when needed. On a laptop CPU, an audit takes
\numRunGlobalAlarm{}--\numRunGlobalIns{} seconds for detection and \numRunFullAlarm{}--\numRunFullIns{}
seconds with every channel diagnosed, model queries included and excluding the one-time fitting of
propagation mechanisms, so it can run on every model update in a release or CI pipeline.

\paraheading{Smaller budgets}
A practitioner with a limited allowance needs to know what a smaller budget costs. We cap queries
from 2{,}500 to 100{,}000 on feature-shuffling candidates; at each cap, each method uses the largest
predeclared test set that fits, chosen without regard to accuracy
(Figure~\ref{fig:accuracy-vs-budget}). \tool{} degrades gracefully: at \numBudgetCap{} queries it
ranks at \numBudgetRankIns{} and \numBudgetRankAlarm{}, against \numShuffleRankIns{} and
\numShuffleRankAlarm{} for a full audit, and it recovers full accuracy by 100{,}000. It leads every
comparator at every cap, including statistical parity difference, the cheapest method at
\numBudgetParityQ{} queries, whose accuracy is fixed because it already uses every test instance;
at \numBudgetCap{} queries \tool{} leads it by \numParityGapCapIns{}--\numParityGapCapAlarm{}.
KernelSHAP first fits at the 100{,}000-query cap, where it reaches \numBudgetShapCap{}.

\phantomsection\finding{
A full \tool{} audit costs 80--400$\times$ fewer queries than SHAP needs for the same test instances
and runs in under a second; \tool{} ranks most accurately at every query budget, and omitting channel
diagnostics cuts network-wide cost by \numSaveIns{} (Insurance) and \numSaveAlarm{} (ALARM).}

\section{Discussion}
\label{sec:discussion}

\tool{}'s central design choice is to treat the causal structure an unlearning audit needs as
part of the test specification, rather than something the testing algorithm must discover. This
follows specification-based testing, where the practitioner states the specification
and the test checks the system against it. This design makes the causal assumptions explicit
and inspectable: the verdict rests on an assumption a practitioner can read and dispute, rather than one
hidden in the choice of check, and because the causal specification names the channels, a failed audit
shows where influence persists, not only that it does. Combined with \tool{}'s prediction-only
probes, it also permits post-deployment auditing, for example as a compliance check in a release or CI pipeline.

\paraheading{Why a Declared Graph}
\emph{The graph is a specification, not something \tool{} claims to infer.} Separating residual
causal influence from association requires causal assumptions, and no test that sees only
predictions and observational data avoids them. The checks we compare against make them
implicitly: a direct-input probe behaves as if the target had no descendants, and a disparity
measure as if every association were influence. \tool{} states its assumptions where a practitioner can
inspect them.
\emph{The required structure is local.} \tool{} needs the target's downstream inputs and their
parents, not a complete causal model, and even that need not be exact: on Adult, a
coarse ordering of the variables sufficed for detection (Section~\ref{sec:robustness-eval}).
\emph{Much of it is already known.} For derived features, the most direct carriers of residual
influence, the downstream relations follow from the feature pipeline that constructs them rather
than from causal discovery, whose graphs on software-analytics data are known to be
unstable~\cite{hulse2025shaky}.

\paraheading{Writing the Graph}
Because over-declared edges failed safe in our experiments (Section~\ref{sec:robustness-eval}), a practitioner unsure of an edge may declare it conservatively: an unneeded edge raised alarms but never hid residual influence, and
the alarm names the channel through that edge. The practitioner can then test the blamed edges against
the data \tool{} already fits on. Such a test cannot detect a reversed edge, and a weak true edge
can look unsupported, so it should prompt review, not dismissal.

\paraheading{Threats to Validity}
\emph{Specification validity.} \tool{}'s causal reading is conditional on the declared structure and mechanisms, and sparse support for intervened parent values can make propagation extrapolative. Results are relative to the declared graph, the test instances, and the contrast; a causal path absent from
the graph is not tested. The bootstrap interval covers variation across test instances, not
uncertainty in the graph or the fitted mechanisms. We have not tested adversarially chosen
graph errors or noninvertible mechanisms; our misspecified mediators are a function of their
parents plus noise.
\emph{Evaluation scope.} Reference effects come from published benchmark mechanisms, controlled
SCMs, a controlled image renderer, and one externally released contrast (IncomeSCM). The
real-data audit has clean controls but no reference effects. The benchmark experiments use one target per network and an MLP on binary outcomes; the
Adult audit adds a random forest and the image study a CNN.
\emph{Deployment conditions.} Strict detection ($\delta=0$) is intentionally stringent: it fires
on negligible influence and, on a noisy prediction interface, needs cached queries or a tolerance
$\delta>0$, a choice \tool{} leaves to the test specification.

\paraheading{Beyond Feature Columns}
The specification-based framing extends past \tool{}'s feature-level test oracle in two
directions. For concept-level targets, the image study shows the approach carries over when the
target is not a column: one predefined binary concept (hospital identity), with a mechanism that
rerenders the same anatomy in another scanner style. Specifying and validating such edits for
broader concepts is the next step. For language models, a concept such as a person or attribute
would be intervened on through counterfactual edits of the input text, and the difficulty is specifying which downstream content an edit should change; adapting paired interventions to text remains open. Record-level targets would intervene on training-set membership.

\section{Related Work}
\label{sec:related}

\medskip\noindent\textbf{Record-Level Unlearning Verification.}
Most work on verifying unlearning targets training records: whether specific examples have
been removed, or whether the model behaves as if it had been retrained without them.
Influence functions and gradient-trajectory analysis estimate a training example's
contribution from parameter gradients~\cite{koh2017understanding,garima2020estimating}, which
requires white-box access. Approximate deletion~\cite{izzo2021approximatedatadeletionmachine}
and memorization analysis~\cite{Jagielski2023MeasuringFO} ask whether traces of deleted samples
remain. Attack-based approaches test forgetting adversarially: dataset updates and unlearning
itself can leak training data~\cite{salem2020updates,chen2021when}, membership inference can test
whether a record was removed~\cite{shokri2017membership,salem2018ml}, Athena plants poisoned
data so that removal can later be checked~\cite{sommer2022athena}, and Goel et
al.~\cite{goel2023towards} evaluate unlearning under realistic, imperfect conditions. This
literature addresses the right problem with the wrong unit of evidence for a feature-level test
oracle: the record, often examined through shadow models, training-time instrumentation, or
specially prepared examples. None of it detects a feature's influence that survives in variables
downstream of that feature.

\medskip\noindent\textbf{Feature Unlearning, Attribution, and Fairness Testing.}
Feature unlearning methods use first-order parameter updates~\cite{warnecke2023unlearning},
mutual-information-based detachment~\cite{guo2022efficient}, or fine-tuning on shuffled
data~\cite{yang2025featureunlearning}, and judge success by comparison with retraining, by
Shapley-value attribution, or by how much information about the feature remains recoverable.
Attribution methods such as SHAP~\cite{lundberg2017unified}, permutation
importance~\cite{fisher2019modelswrongusefullearning}, and
saliency~\cite{simonyan2014deepinsideconvolutionalnetworks} measure predictive dependence, and
have been applied to verifying unlearning~\cite{vidal2025verifying}. Fairness
testing~\cite{brun2018softwarefairness,chen2024fairness}, the nearest methodology to ours, tests a
declared sensitive
attribute~\cite{galhotra2017themis,tramer2017fairtest,aggarwal2019bbft,black2020fliptest,bellamy2018aifairness360extensible}
or searches for discriminatory inputs
anywhere~\cite{udeshi2018automated,zhang2020whitebox,zheng2022neuronfair,zhang2021gradsearch,fan2022explanation,xiao2023latentimitator,monjezi2025evtfair},
whereas an unlearning audit measures one feature's influence on one deployed model. Bias
mitigation work has long observed that removing a sensitive attribute from the inputs leaves its
influence in correlated proxies~\cite{biasmitigation2024survey}; \tool{} turns this observation
into a test oracle. None of these
checks suffices as a feature-level test oracle: retraining leaves influence in the feature's
descendants, perturbation sees nothing once the feature is gone from the inputs, group
comparisons conflate influence with correlation, and recoverable information need not be influence.

\medskip\noindent\textbf{Causal and Specification-Based Testing.}
Causal fairness supplies the relevant effect notions. Path-specific and counterfactual
formulations separate direct, indirect, and merely correlated influence of a sensitive
attribute~\cite{zhang2019causal,nabi2018se,kusner2017counterfactual,silvia2019pathspecific}.
We build on these notions but apply them to a different problem: this line of work mostly
constrains how a predictor is designed or trained under an assumed causal model, and treats some
pathways as permissible. \tool{} instead \emph{tests} an already-deployed model through its
prediction interface alone, and treats every causal pathway from the erased feature as
prohibited. Closest in spirit is CausalFT~\cite{du2026causally}, which builds a causal DAG from
data to guide fairness test generation, but extracts only \emph{direct} edges from the sensitive
attribute to the outcome and uses them to guide search rather than to measure the mediated
influence an unlearning request needs erased. A causal graph alone is also not enough: backdoor
adjustment, given the same graph, misses most failures that paired interventions catch
(Section~\ref{sec:detection-eval}).

\tool{} also fits the tradition of specification-based and metamorphic
testing~\cite{segura2016survey}, where correctness is judged against a test oracle stated outside
the system under test~\cite{barr2015oracle}. In \tool{}'s feature-level instantiation, causal
non-influence within the declared tolerance is the test oracle; the causal specification
determines how intervened inputs are constructed, and the paired interventions instantiate the
test. Metamorphic relations are already used to test ML systems for which no per-input expected
output exists~\cite{zhang2022mltesting,tian2018deeptest}; \tool{} applies them to unlearning.
Mazhar et al.~\cite{mazhar2026towards} argue for target-specific test oracles and outline
causal testing for features; \tool{} realizes such a test.

\section{Conclusion}
\label{sec:conclusion}

As ML models become software components that must adapt to changing requirements, unlearning must
be tested like any other update, and for features a model's inputs are not enough: a feature's
influence can outlive its removal. \tool{} tests feature unlearning for causal non-influence through
paired interventions on a deployed model's predictions. In our evaluation, it exposed
failures that direct-input checks certified as forgotten, without false alarms on clean controls,
and localized where residual influence persists.

\clearpage
\bibliographystyle{ACM-Reference-Format}
\bibliography{ref}
\end{document}